\documentclass[sigconf]{acmart}
\AtBeginDocument{%
  }

\copyrightyear{2025}
\acmYear{2025}
\setcopyright{acmlicensed}\acmConference[MM '25]{Proceedings of the 33rd ACM International Conference on Multimedia}{October 27--31, 2025}{Dublin,Ireland}
\acmBooktitle{Proceedings of the 33rd ACM International Conference on Multimedia (MM '25), October 27--31, 2025, Dublin, Ireland}
\acmDOI{10.1145/3746027.3755598}
\acmISBN{979-8-4007-2035-2/2025/10}

\usepackage{booktabs}
\usepackage{makecell}
\usepackage{multirow}
\usepackage{graphicx}
\usepackage{subfigure}
\usepackage{colortbl}

\begin{document}

\title{EHVC: Efficient Hierarchical Reference and Quality Structure for Neural Video Coding}

\author{Junqi Liao}
\authornote{Work done during internship at Bytedance China.}
\orcid{0009-0006-4030-174X}
\email{liaojq@mail.ustc.edu.cn}
\affiliation{%
  \institution{University of Science and Technology of China}
  \city{Hefei}
  \state{Anhui Province}
  \country{China}
}

\author{Yaojun Wu}
\orcid{0000-0002-8138-4186}
\email{wuyaojun@bytedance.com}
\affiliation{%
  \institution{Bytedance China}
  \city{Beijing}
  \country{China}
}

\author{Chaoyi Lin}
\orcid{0009-0002-7770-6821}
\email{linchaoyi.cy@bytedance.com}
\affiliation{%
  \institution{Bytedance China}
  \city{Hangzhou}
  \state{Zhejiang Province}
  \country{China}
}

\author{Zhipin Deng}
\orcid{0009-0007-9854-9470}
\email{zhipin.deng@bytedance.com}
\affiliation{%
  \institution{Bytedance China}
  \city{Beijing}
  \country{China}
}

\author{Li Li}
\authornote{Corresponding author}
\orcid{0000-0002-7163-6263}
\email{lil1@ustc.edu.cn}
\affiliation{%
  \institution{University of Science and Technology of China}
  \city{Hefei}
  \state{Anhui Province}
  \country{China}
}

\author{Dong Liu}
\orcid{0000-0001-9100-2906}
\email{dongeliu@ustc.edu.cn}
\author{Xiaoyan Sun}
\orcid{0000-0003-3638-5566}
\email{sunxiaoyan@ustc.edu.cn}
\affiliation{%
  \institution{University of Science and Technology of China}
  \city{Hefei}
  \state{Anhui Province}
  \country{China}
}

\renewcommand{\shortauthors}{Junqi Liao et al.}
\begin{abstract}
  Neural video codecs (NVCs), leveraging the power of end-to-end learning, have demonstrated remarkable coding efficiency improvements over traditional video codecs. Recent research has begun to pay attention to the quality structures in NVCs, optimizing them by introducing explicit hierarchical designs. However, less attention has been paid to the reference structure design, which fundamentally should be aligned with the hierarchical quality structure. In addition, there is still significant room for further optimization of the hierarchical quality structure. To address these challenges in NVCs, we propose EHVC, an efficient hierarchical neural video codec featuring three key innovations: (1) a hierarchical multi-reference scheme that draws on traditional video codec design to align reference and quality structures, thereby addressing the reference-quality mismatch; (2) a lookahead strategy to utilize an encoder-side context from future frames to enhance the quality structure; (3) a layer-wise quality scale with random quality training strategy to stabilize quality structures during inference. With these improvements, EHVC achieves significantly superior performance to the state-of-the-art NVCs. Code will be released in: https://github.com/bytedance/NEVC.
  \vspace{-0.8em}
\end{abstract}


\vspace{-0.8em}

\maketitle

\begin{figure}[t]
  \centering
  \includegraphics[width=8.0cm]{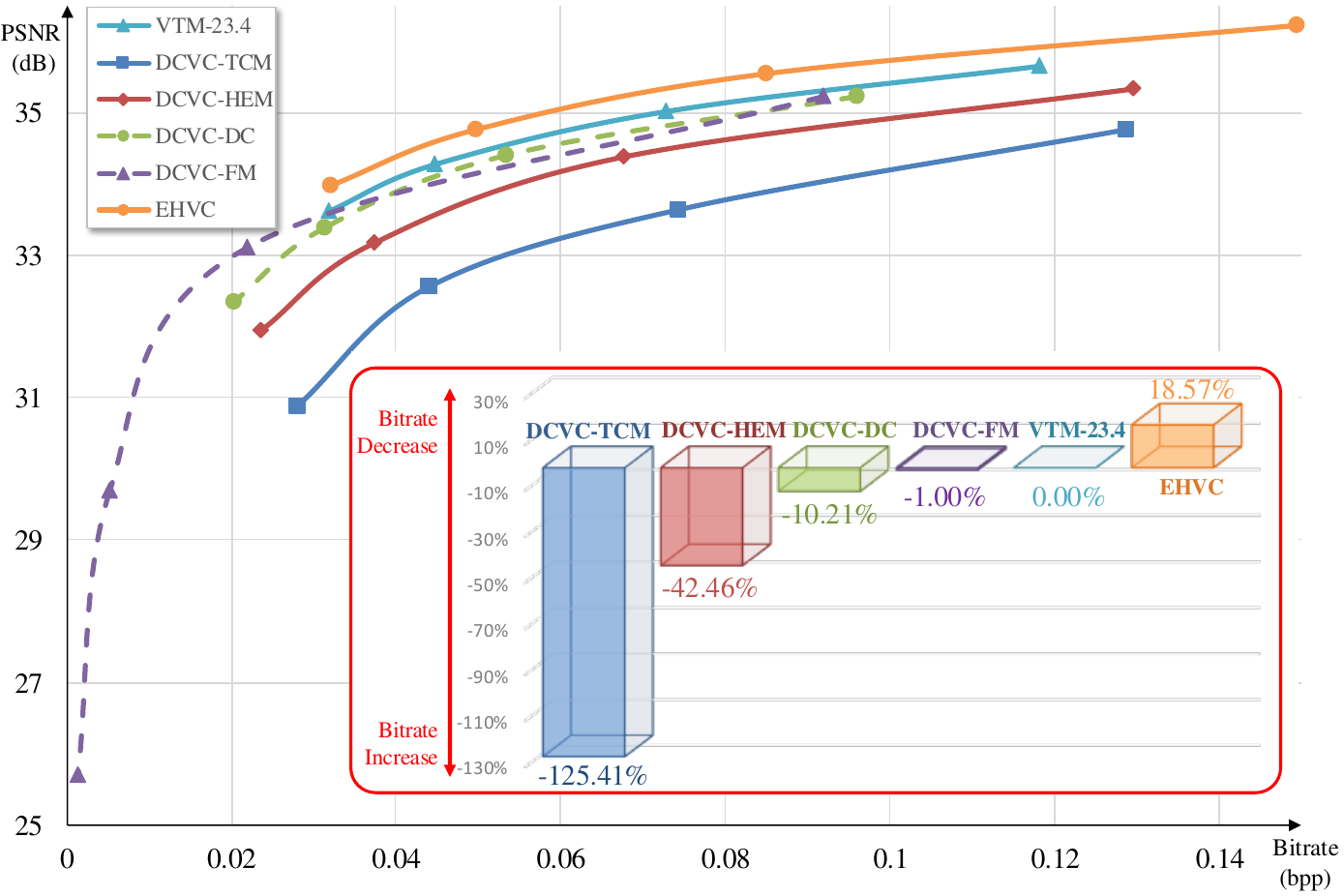}
  \vspace{-1.0em}
  \caption{Rate-Distortion curves and BD-rate~\cite{bjontegaard2001calculation} comparisons with other video codecs. The test dataset is HEVC B with intra-period -1. EHVC surpasses VTM and SOTA NVC.}
  \label{fig-performance}
\vspace{-0.8em}
\end{figure}

\vspace{-0.8em}
\section{Introduction}
Traditional video codecs with the hybrid residual coding framework have been mainstream from H.261~\cite{girod1995comparison} in 1988 to H.266~\cite{bross2021overview} in 2020. Despite their evolution, they face inherent limitations: complexity~\cite{bossen2021vvc} and interdependent coding tools~\cite{mercat2021comparative,bross2021developments}, hindering further efficiency gains. Meanwhile, the recent breakthroughs in deep neural networks have catalyzed a paradigm shift in video compression, sparking research interest in neural video codecs (NVCs) as they demonstrate the potential to break the performance ceilings of traditional coding approaches.

The NVCs can be broadly categorized into two distinct paradigms based on their fundamental coding frameworks: residual and conditional coding-based NVCs. The early NVCs adopt the residual coding framework inherited from traditional video codecs. This paradigm performs motion compensation directly in the pixel domain to generate predictions, followed by residual coding. However, more recent advances have shifted toward the conditional coding framework~\cite{ho2022canf,ladune2021conditional,liu2020conditional,mentzer2022vct,qi2023motion,guo2023enhanced,xiang2022mimt,chen2023neural,li2021deep,sheng2022temporal,li2022hybrid,li2023neural,li2024neural}, which offers superior flexibility and efficiency. Unlike pixel-domain residual coding, conditional coding operates entirely in the feature domain, enabling more sophisticated information representation and processing. This architectural advantage, combined with recent innovations in module design and training methodologies, has allowed state-of-the-art conditional coding-based NVCs like DCVC-DC~\cite{li2023neural} and DCVC-FM~\cite{li2024neural} to demonstrate remarkable performance gains, even surpassing the coding efficiency of VTM (the reference software of H.266/VVC) under specific configurations.

While significant research efforts have been devoted to optimizing neural network architectures and coding modules in NVCs, the top-level design of NVC such as hierarchical structure (including quality structure and reference structure) remains largely unexplored. This stands in contrast to the traditional video codecs, where decades of development have optimized not only module-level algorithms but also the overarching hierarchical structure - a key factor enabling their robust and efficient performance~\cite{wang2021high}. For example, in VTM, the different quantization parameter offsets and reference picture lists for frames in different layers define the hierarchical quality structure and reference structure, respectively. These two carefully designed efficient hierarchical structures work together to enable traditional video codecs to achieve stable and efficient performance when dealing with videos with different content.

Unlike the traditional video codecs that explicitly constrain the hand-designed hierarchical structure, NVCs typically exhibit implicit hierarchical structures through their network architectures. This fundamental difference is reflected in two key aspects: quality structure and reference structure. For the quality structure, state-of-the-art (SOTA) NVCs employ lossy neural networks as their coding backbone, where quality degradation emerges as an inherent property of the network's nonlinear transformations rather than being explicitly controlled. For reference structure, SOTA NVCs use the conditional coding-based framework where inter-frame information is implicitly passed in the feature domain, resulting in an implicit reference structure. 

Theoretically, a sufficiently trained NVC should be capable of learning the optimal implicit hierarchical structure. However, practical constraints on training resources limit the multi-frame training stage to short video sequences, which creates a critical mismatch between training and inference conditions. This limitation prevents the learned implicit hierarchical structure from effectively generalizing to variable-length sequences during inference. More importantly, in scenarios involving long prediction chains, such train-test mismatch can lead to significant performance degradation and even cause error propagation, substantially compromising the codec's practical applicability~\cite{li2024neural}.

The hierarchical structure design initially received limited attention in early NVC research, primarily because these works typically evaluated performance only under intra-period 32 configuration. In such constrained settings, the implicitly learned hierarchical structures from finite-length training sequences proved sufficient for handling short prediction chains. However, as research advanced, the limitations of purely implicit structural learning became apparent, prompting systematic efforts to optimize hierarchical structure. DCVC-DC~\cite{li2023neural} proposes to learn from traditional video codecs' hierarchical quality structure design and introduces hierarchical lambda weights [0.5, 1.2, 0.5, 0.9] for NVC training to guide the implicit hierarchical structure. DCVC-FM~\cite{li2024neural} first identifies the error propagation problem of NVC and tries to mitigate it by training on longer sequences and pulling up the reconstructed frame quality by periodically refreshing context generation. 


Although recent studies have begun optimizing the hierarchical quality structure in NVCs, two unsolved challenges remain. The primary challenge lies in the reference-quality mismatch. Traditional video codecs maintain strict correspondence between reference structure and quality structure. This is because high-quality frames have a higher reference value and should have a greater probability of being referenced by subsequent frames. In contrast, the existing NVCs improving hierarchical structure ignore this point and only focus on optimizing the quality structure, using purely implicit reference structures. Another challenge is that the hierarchical quality structure is still inefficient. There is much room to improve the efficiency and stability of the hierarchical quality structure in NVC.

In this work, by considering the reference-quality mismatch and the inefficiencies in the existing quality structure, we propose an NVC with a more efficient hierarchical structure, EHVC. Specifically, we first design a hierarchical reference structure that draws on the design in traditional video codecs to align reference and quality structures. To achieve the designed reference structure, we introduce a hierarchical multi-reference scheme. Then, we propose a lookahead strategy using forward features to enrich the encoder-side context generation and improve the quality structure. In addition, we introduce the layer-wise quality scale with a random quality training strategy to improve the stability and flexibility of the quality structure. As shown in Fig.~\ref{fig-performance}, our EHVC not only outperforms SOTA traditional video codec VTM~\cite{bossen2021vvc} but also surpasses SOTA NVC DCVC-FM~\cite{li2024neural} in rate-distortion performance. Our contributions can be summarized as follows:

\begin{itemize}
\item We design a hierarchical reference structure for NVC in low delay configuration. We propose a corresponding hierarchical multi-reference scheme, aligning reference and quality structures and addressing reference-quality mismatch.

\item We propose the lookahead strategy, enriching the encoder-side context with forward features and achieving better rate-distortion performance.

\item We propose the layer-wise quantization scale with a random quality training strategy, making it possible to learn a more flexible and efficient quality structure.

\item Our EHVC can outperform SOTA traditional video codec VTM under intra-period -1 setting. Our EHVC shows a better hierarchical structure than previous NVCs. In addition, our EHVC reduces 10.98\% and 12.88\% more bitrates than previous SOTA NVC DCVC-FM over VTM-23.4 low-delay B (LDB) under the intra-period of 32 and -1, respectively.
\end{itemize}

\vspace{-0.8em}
\section{Related Work}
\subsection{Neural Video Coding}
Existed NVCs can be intuitively categorized into two main types: residual coding-based NVCs and conditional coding-based NVCs. Early NVCs inherit the idea of the traditional video codecs using the framework based on residual coding. These NVCs get the prediction frame through pixel-domain motion compensation and encode the motion and residual data. The pioneer in this category is DVC~\cite{lu2019dvc}, which utilizes convolutional neural networks to construct an end-to-end residual coding structure. Following this paradigm, many NVCs~\cite{agustsson2020scale,djelouah2019neural,hu2022coarse,lin2020m,liu2020neural,lu2020end,rippel2021elf,hu2020improving} enhance this framework by designing stronger sub-modules and sophisticated approaches. For example, a coarse-to-fine motion coding~\cite{hu2022coarse} is proposed to enhance the motion accuracy. The multi-scale motion compensation and adaptive spatiotemporal context model are introduced~\cite{liu2020neural} to improve the inter-frame information coding. A flexible-rate framework and an in-loop flow prediction scheme are proposed~\cite{rippel2021elf} to improve the bitrate flexibility and compression efficiency.

In contrast to using pixel-domain prediction as the temporal context, conditional coding-based NVCs use the feature-domain temporal context. The conditional coding framework is more flexible and has higher potential since the context is not limited to the pixel-domain predicted frame and the way to remove redundancy is no longer limited to the subtraction. Some early works~\cite{habibian2019video,lombardo2019deep} use the latent feature extracted from all previous compressed frames as the temporal context. In contrast, the DCVC family~\cite{li2021deep, sheng2022temporal, li2022hybrid, li2023neural, li2024neural} propose using feature-domain motion compensation to extract the temporal context consecutively. Many modern NVCs~\cite{tang2025neural,bian2025augmented,liao2025efvc} follow this paradigm. Recent DCVC-FM~\cite{li2024neural} has demonstrated performance beyond the latest traditional video coding standard H.266/VVC~\cite{bross2021overview} under intra-period -1 setting.

Though the above NVCs have achieved notable coding performance through model design and training optimization, many problems at the top level above the modules are not taken seriously. The design of the hierarchical structure (including the quality and reference structures) is an important problem. For traditional video codecs, the hand-crafted hierarchical structure is usually optimized with the iteration of the codec to achieve more stable and efficient coding quality. For NVCs, an efficient hierarchical structure becomes a goal of end-to-end learning. However, inadequate training makes the learned hierarchical structure suboptimal. Therefore, this work introduces an efficient hierarchical NVC named EHVC, optimizing the hierarchical reference and quality structures.

\vspace{-0.8em}
\subsection{Hierarchical Structure Optimization}
\subsubsection{Quality structure optimization}
The enhancement of the quality structure optimizes the frame-level bit allocation. Most NVCs learn the quality structure through multi-frame training instead of explicitly optimizing. A frame-level bit allocation network~\cite{zhang2023neural} is designed to obtain the current frame's target bitrate according to all frames in the sequence and the remaining bitrate budget. However, this work is in the rate control scenario and does not optimize the quality structure in the general scenario. The hierarchical lambda weights~\cite{li2023neural} ([0.5,1.2,0.5,0.9]) are proposed to optimize the quality structure in DCVC-DC. The weights divide frames into three layers to form a periodic group-of-pictures (GOP) of length four. The instability of the implicit quality structure is obvious in a long prediction chain, which leads to error propagation. The context generation periodical refresh~\cite{li2024neural} is proposed to pull up the reconstruction quality and inhibit error propagation periodically.

\subsubsection{Reference structure optimization}
Many works~\cite{lin2020m,park2019deep,yang2020learning,park2020deep,yilmaz2020end, sheng2025bi,alexandre2023hierarchical} try to improve the reference structure of NVC. These works can be divided into low delay structure (LD) and random access structure (RA) according to whether the current frame refers to the frame following it in the play order. In the first category, multi-frame prediction is introduced to improve the coding efficiency~\cite{lin2020m}. However, this work only introduces t-2 frame as the reference and does not design the reference structure carefully. In the second category, based on the excellent performance of random access in traditional coding, many attempts~\cite{park2019deep,yang2020learning,park2020deep,yilmaz2020end,sheng2025bi,alexandre2023hierarchical} are made to introduce it in NVC. However, these attempts on RA have not demonstrated significant performance gain over SOTA NVC in LD.

Though these attempts to improve NVC's hierarchical structure have achieved some performance, there are still obvious problems. First, existing NVCs do not consider designing the reference structure corresponding to the hierarchical quality structure. In addition, most of these attempts to improve quality structures are not carefully designed and do not fully learn from the mature hierarchical structure of traditional video coding. To address these problems, this work draws on the hierarchical structure design in traditional video coding, optimizes the reference structure and quality structure in NVC, and finally achieves performance beyond SOTA NVCs.

\vspace{-0.8em}
\section{Analysis}

\subsection{Quality structure}

\begin{figure}[t]
  \centering
  \includegraphics[width=7.5cm]{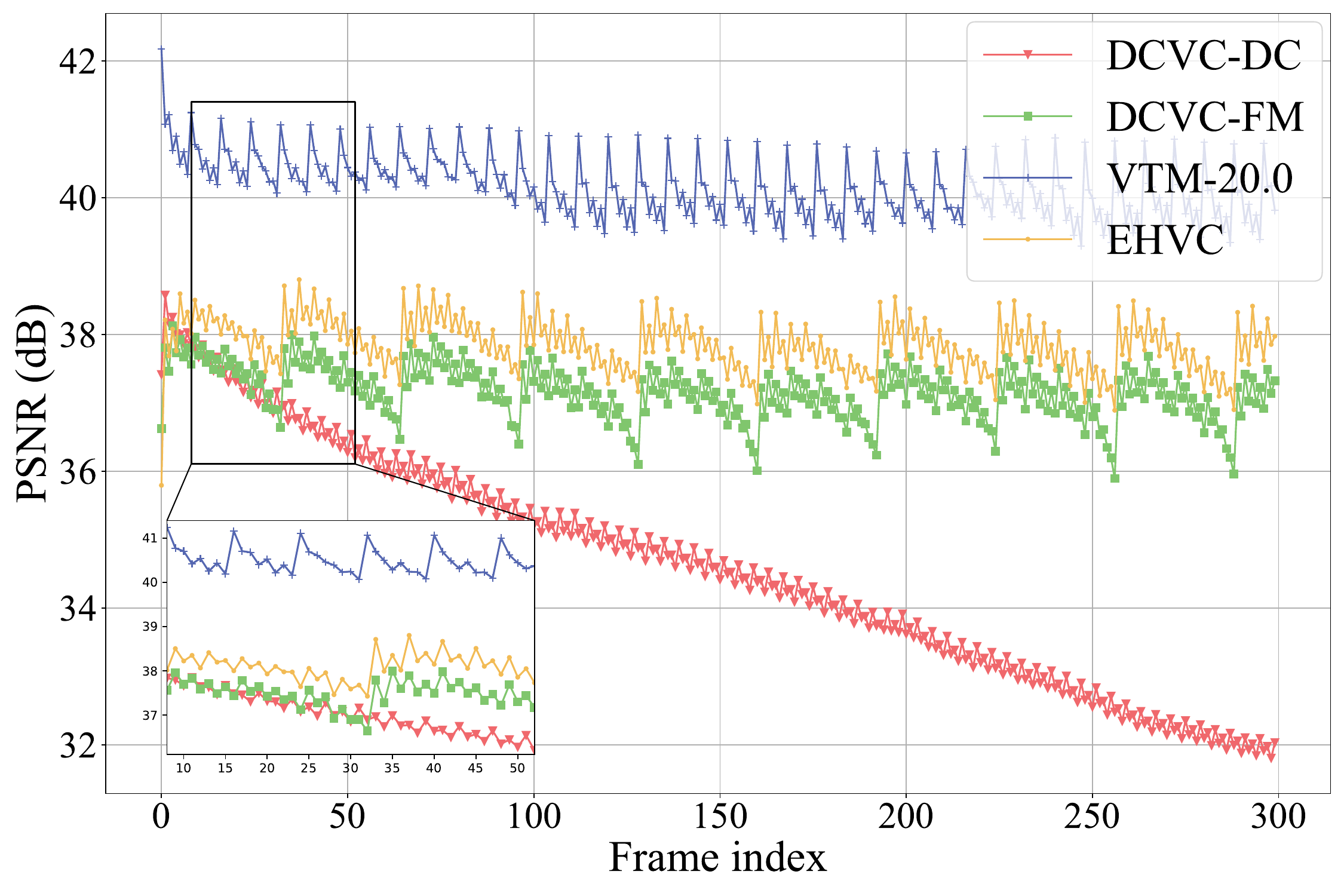}
  \caption{Frame-level PSNR on the first 300 frames of FourPeople using different codecs with intra-period -1. These codecs' average bitrates (bits per pixel) are aligned at 0.01.}
  \label{fig-framePSNR}
  \vspace{-1.5em}
\end{figure}
Using the first 300 frames of the FourPeople sequence (a standard test sequence in VTM common test conditions) as our test case, Fig.~\ref{fig-framePSNR} demonstrates the quality fluctuations across various state-of-the-art video codecs. All evaluated codecs employ an intra-period of -1 (only the first frame is intra-coded), and VTM-20.0 is under low-delay B configuration.

\subsubsection{Quality structure in SOTA traditional video codec}
As illustrated in Fig.~\ref{fig-framePSNR}, VTM employs an 8-frame cyclic quality structure, corresponding to its GOP size 8 in low-delay B configuration. The quantization parameter (QP) for the i-th frame within a GOP is determined by:
\begin{equation}
\label{eq:QPcalc}
QP_i = Int\left(QP_{bias,i}+MScale_iQP_{bias,i}+MOffset_i+0.5\right),
\end{equation}
where $Int\left(\cdot\right)$ denotes the truncation operation that discards the fractional part of a real number, $QP_{bias,i} = QP_{base}+Offset_i$, the $QP_{base}$ is the predefined base QP determining the overall quality level of the reconstructed sequence. The parameters, including $Offset$, $MScale$, and $MOffest$, are obtained according to the frame index in GOP and the correspondence shown in Table~\ref{tab:QPcalc}. As shown in the table, frames are organized into three hierarchical layers within each GOP, with an alternating pattern of high- and low-quality frames strategically distributed throughout the group.

\begin{table}[t]
\caption{The QP calculation parameters in VTM under LDB configuration\label{tab:QPcalc}}
\centering
\begin{tabular}{ccccc}
\toprule
\begin{tabular}[c]{@{}c@{}}Frame index \\ in GOP\end{tabular}&$Offset$&$MScale$&$MOffset$&Quality \\
\midrule
0, 2, 4, 6  &6  &0.245  &-6.5 &low \\
1, 3, 5     &4  &0.259  &-6.5 &high\\
7           &1  &0.0    &0.0  &very high\\
\bottomrule
\end{tabular}
\vspace{-1.2em}
\end{table}

Since the QPs directly determine the quality loss in the quantization, and the relationship between the QP and the Lagrange multiplier $\lambda$ in the rate-distortion optimization (RDO) objective is determined, the quality structure in VTM is completely \textbf{explicit}.

\subsubsection{Quality structure in SOTA NVC}
As shown in Fig.~\ref{fig-framePSNR}, both DCVC-DC and DCVC-FM adopt a quality structure with alternating high- and low-quality frames. This design stems from their shared use of hierarchical Lagrange multiplier weights ([0.5, 1.2, 0.5, 0.9]), initially proposed in DCVC-DC. During multi-frame training, the weighted loss function can be expressed as:
\begin{equation}
\label{eq:lossFunc}
L = \sum_{i=1}^{N}{\omega_{i\%4}\lambda D_i+R_i},
\end{equation}
where $N$ is the number of inter-coded frames, the $D_i$ and $R_i$ are the distortion (MSE) and bitrate (bits per pixel) of the $i$th inter-coded frame. The $\omega_{i\%4}$ is the hierarchical weight of the $i$th inter-coded frame. The $\lambda$ is the predefined Lagrange multiplier that governs the target rate-distortion trade-off during training.

Furthermore, DCVC-DC exhibits a noticeable quality degradation as the frame index increases, a phenomenon known as error propagation in long prediction chains. This issue stems from the instability in its quality structure. To alleviate error propagation, DCVC-FM periodically refreshes the temporal context generation to a high-quality branch to pull up the reconstruction quality. The refresh period is set as 32 frames in DCVC-FM. Therefore, for DCVC-FM, it is evident that there is a quality pull every 32 frames.

In summary, while end-to-end learning represents a significant advantage of NVCs, the learned \textbf{implicit} quality structure often proves inefficient during inference due to insufficient training. This limitation stems primarily from computational constraints that prevent comprehensive multi-frame training on long prediction chains. Consequently, the implicit quality structure fails to generalize well to sequences on long prediction chains during inference - a phenomenon known as train-test mismatch. To address this challenge, both DCVC-DC and DCVC-FM incorporate designed \textbf{explicit} constraints into their quality structures, achieving measurable performance improvements. 

However, current explicit constraints on hierarchical quality structures in NVCs suffer from two limitations. First, the quality structure relies exclusively on backward temporal information without forward information. Second, the design of hierarchical weights only optimizes the quality structure hidden in the model parameters by constraining the training target and lacks the hierarchical design of the quantization parameters. These limitations motivate the innovations proposed in our EHVC.

\vspace{-0.8em}
\subsection{Reference structure}
\begin{figure}[t]
  \centering
  \includegraphics[width=8cm]{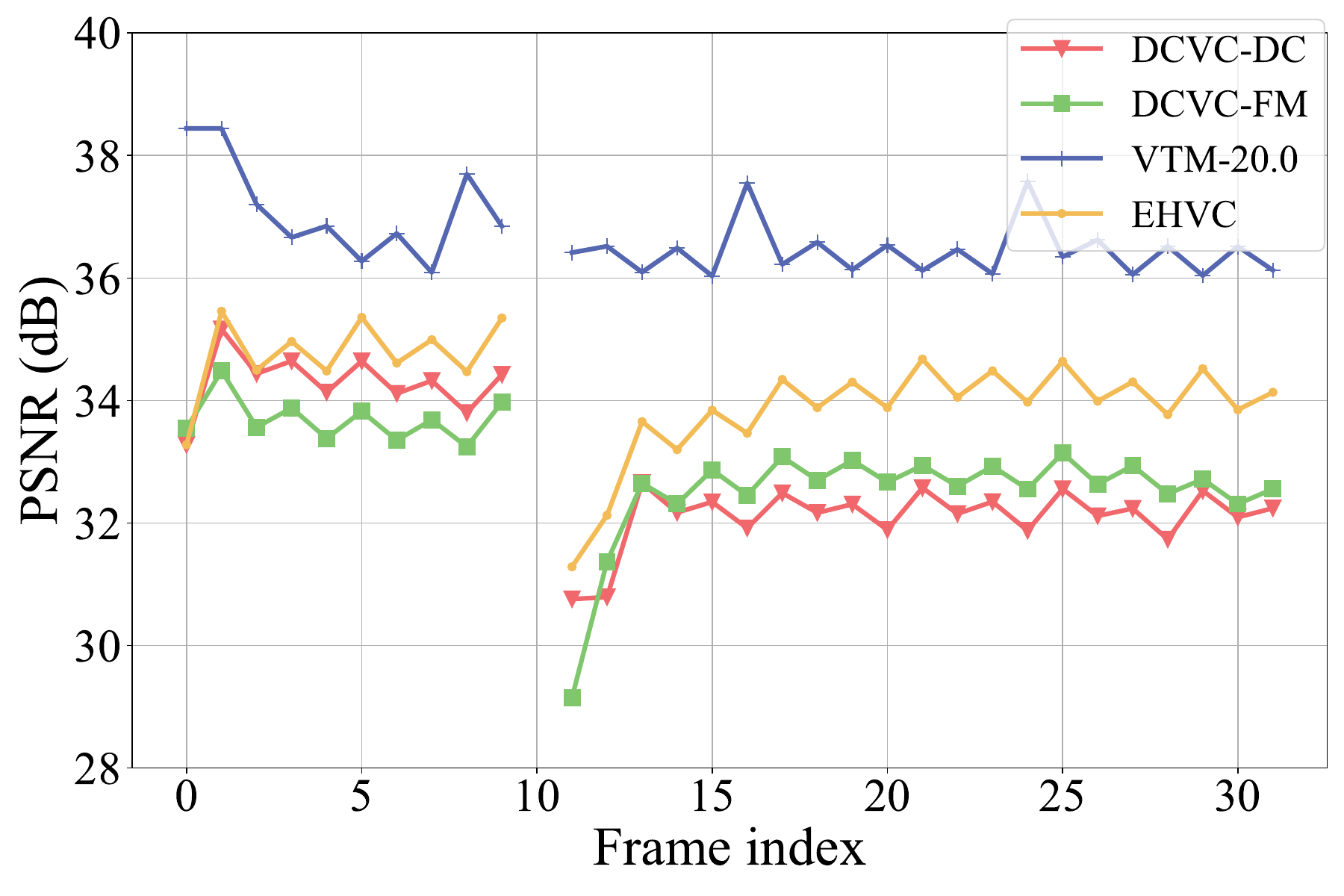}
  \caption{Frame-level PSNR on the first 32 frames of BasketballDrill using different codecs with intra-period -1. To test the reference structure stability, the frame with index 10 is set to a no-information frame (all pixels are set to 128). These codecs' average bitrates (bits per pixel) are aligned at 0.04.}
  \label{fig-blackF}
  \vspace{-1.2em}
\end{figure}
To evaluate the stability of reference structures in video codecs, we test the coding performance of the SOTA traditional video codec and SOTA NVCs in the case where a frame of the sequence provides no reference information. Specifically, the frame with index 10 is set to a no-information frame (all pixels are set to 128) without any valuable reference information. The test results are shown in Fig.~\ref{fig-blackF}. All codecs use intra-period -1 (only the first frame uses intra-coding), and VTM-20.0 is under low-delay B configuration. Since all codecs' PSNRs of the no-information frame are too high, the corresponding points are deleted to make the curves clearer.

\begin{figure*}[t]
  \centering
  \includegraphics[width=11cm]{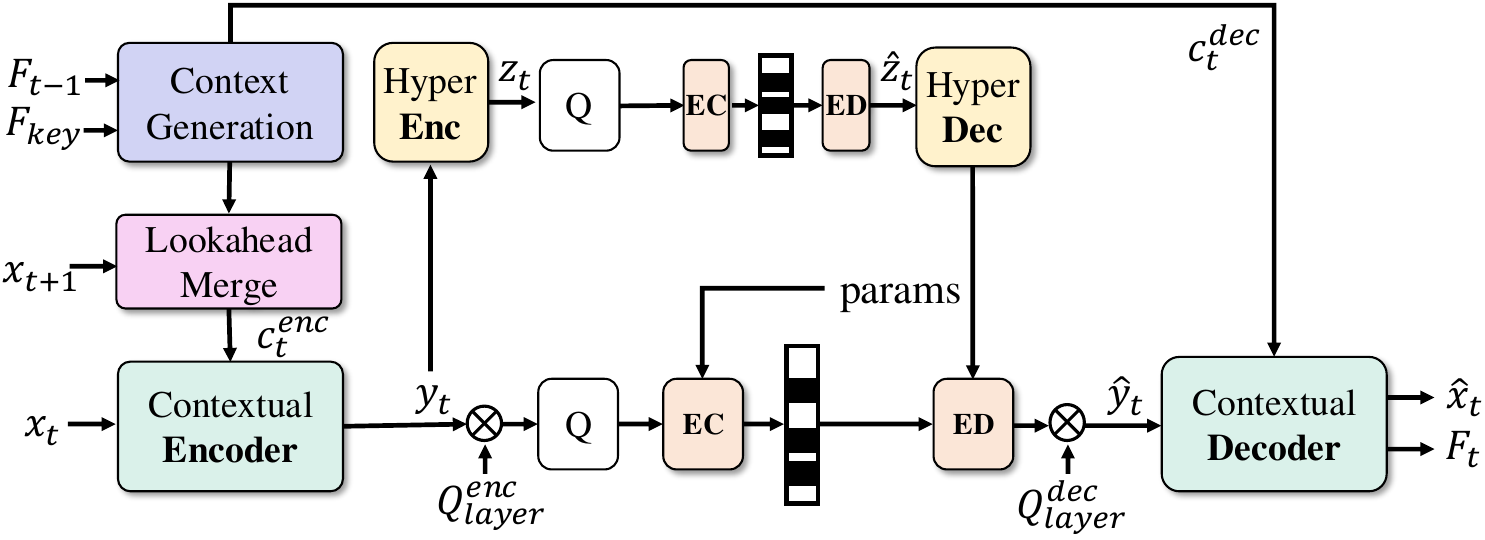}
  \caption{The framework of our EHVC. The motion coding branch follows the same structure as DCVC-FM and is therefore omitted here. The Q denotes the quantization process. The Enc and the Dec denote the encoder and the decoder, respectively. The EC and the ED denote the entropy encoder and the entropy decoder, respectively. }
  \label{fig-framework}
  \vspace{-1.2em}
\end{figure*}

\subsubsection{Reference structure in SOTA traditional video codec}
\begin{table}[t]
\caption{The reference lists in VTM under LDB configuration\label{tab:referVTM}}
\centering
\begin{tabular}{ccccc}
\toprule
\begin{tabular}[c]{@{}c@{}}Frame index \\ in GOP\end{tabular}& Quality & Reference list 0 & Reference list 1 \\
\midrule
0   & low  &1, 9, 17, 25  &1, 3, 5, 33  \\
1   & high &1, 2, 10, 18  &1, 2, 4, 26  \\
2   & low  &1, 3, 11, 19  &1, 3, 5, 27  \\
3   & high &1, 4, 12, 20  &1, 2, 4, 28  \\
4   & low  &1, 5, 13, 21  &1, 3, 5, 29  \\
5   & high &1, 6, 14, 22  &1, 2, 6, 30  \\
6   & low  &1, 7, 15, 23  &1, 3, 7, 31  \\
7   & very high &1, 8, 16, 24  &1, 2, 4, 32  \\
\bottomrule
\end{tabular}
\vspace{-1.4em}
\end{table}
For low-delay B configuration in VTM, there are two reference lists for each frame. Frames can select a reference frame from each list, then fuse the corresponding predictions to get the final prediction. The reference lists are set as shown in Table~\ref{tab:referVTM}. From the table, the numbers in the reference list indicate the distances of the reference frame from the current frame, and a positive number indicates that the reference frame is before the current frame. Therefore, it can be found that the first reference frames of both lists are the neighboring previous frame, and the other reference frames are the nearest high- or very-high quality frames. This is because the previous frame can provide the most similar content to the current frame as a reference, and high-quality reference can provide high-quality information to improve the reconstruction. Therefore, the explicit reference structure in VTM is coupled with its quality structure.

From Fig.~\ref{fig-blackF}, the explicit reference structure in VTM is stable. The subsequent frames of the no-information frame immediately return to the normal quality and are barely affected.

\vspace{-0.7em}
\subsubsection{Reference structure in SOTA NVC}

The conditional coding framework has become a prevalent architecture in SOTA NVCs, where inter-frame dependencies are modeled through feature-domain temporal contexts. These contexts are generated via a series of operations, including motion compensation, information extraction, and multi-scale fusion, which are applied to the previous frame's latent representation. While this feature-domain approach provides greater flexibility than traditional pixel-domain prediction, the implicit nature of these contexts complicates reference structure analysis. From a theoretical perspective, since the context is computed recursively, it should inherently incorporate information from all preceding frames (t-1, t-2,...) rather than being limited to only the immediate previous frame (t-1).

However, as shown in Fig.~\ref{fig-blackF}, the implicit reference structure is unstable. The quality of all frames after the no-information frame is maintained at a lower level. This result indicates that the temporal context does not retain enough information from frames before t-1, and the implicit reference structure is fragile.

In summary, compared with the explicit reference structure, the \textbf{implicit} reference structures of NVCs are unreliable. Therefore, the reference structure in NVC also needs to introduce \textbf{explicit} constraints like the quality structure. Moreover, like traditional video coding, the reference structure should correspond to the quality structure. That is, the \textbf{reference-quality} mismatch should be avoided.

\vspace{-0.8em}
\section{Methodology}
\subsection{Overview}
Our EHVC adopts a conditional coding-based framework, which is built on DCVC-FM. The overall framework is illustrated in Fig.~\ref{fig-framework}.

For coding the t-th frame $x_t$, the contextual encoder first transforms it into latent representation $y_t$ using the encoder-side context $c_t^{enc}$. Subsequently, $y_t$ is fed into the hyperprior encoder (Hyper Enc) to get the hyperprior $z_t$. The $z_t$ is quantized and entropy coded (EC) to obtain the hyperprior's bitstream. The $y_t$ is \textbf{hierarchically quantized} (Section 4.4) and then entropy coded (EC) to obtain a bitstream based on the estimated distribution parameters (params) obtained after hyperprior decoding. The decoder side can obtain the temporal feature $f_t$ and the reconstructed frame $\hat{x}_t$ by a process almost symmetric with the encoder side. To get the context for the t-th frame, the decoder-side context $c_t^{dec}$ can be obtained by the \textbf{context generation based on the previous frame's temporal feature $f_{t-1}$ and the key frame's temporal feature $f_{key}$} (Section 4.2). Furthermore, the encoder-side context $c_t^{enc}$ can be obtained by the \textbf{lookahead merge based on the next frame $x_{t+1}$ and the $c_t^{dec}$} (Section 4.3).

\begin{figure*}[t]
  \centering
  \includegraphics[width=16.2cm]{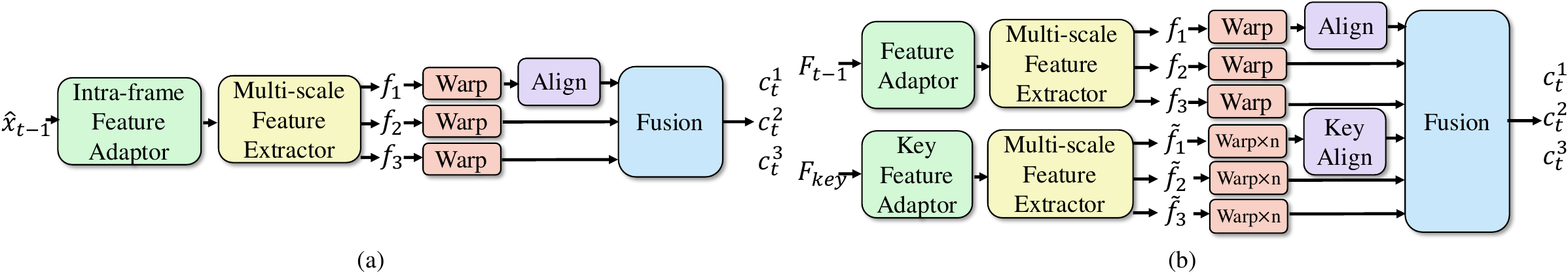}
  \vspace{-1.0em}
  \caption{The context generation process for (a) the intra-coding frame and (b) other frames in our EHVC.}
  \label{fig-contextGen}
  \vspace{-1.0em}
\end{figure*}

\vspace{-1.0em}
\subsection{Hierarchical multi-reference}
\begin{figure}[t]
  \centering
  \includegraphics[width=7.0cm]{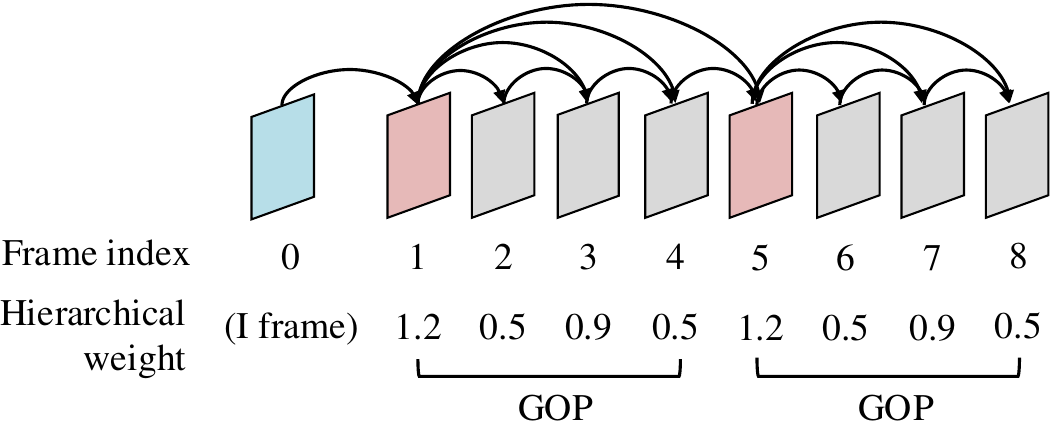}
  \caption{An example of the designed hierarchical reference structure aligned with the quality structure. The blue and red frames are intra-coding frames and key frames, respectively.}
  \label{fig-refStructure}
  \vspace{-1.8em}
\end{figure}

To align the reference structure and the quality structure in NVC, we draw on the design in the traditional video codecs and introduce a hierarchical reference structure in NVC. An example of our designed reference structure on a 9-frame sequence is shown in Fig.~\ref{fig-refStructure}. As shown in the figure, we define the frames with a hierarchical weight of 1.2 (high reconstruction quality) as the key frames (red frames in the figure). Moreover, except for the intra-coding frame and the first key frame, each frame references both its neighboring previous frame and the previous key frame.

Compared to setting the adjacent frame as a single reference, our EHVC maintains it as a reference while incorporating a high-quality key frame reference. This design creates a more stable reference structure. When the adjacent frame is low-quality in the quality structure, the key frame reference can provide good-quality reference information for the current frame.

To realize our hierarchical reference structure, we design the context generation module as shown in Fig.~\ref{fig-contextGen}. Specifically, for the intra-coding frames, the contexts in different scales ($c_t^1$, $c_t^2$, and $c_t^3$) are generated from the reconstructed frame $\hat{x}_t$. The feature adaptor first transforms the $\hat{x}_t$ to the hidden space. Notably, the feature adaptor is a module using different parameters for frames in different layers. Then, the feature is further extracted and separated into three scales (no downsampling, 1x downsampling, and 2x downsampling) to obtain $f_1$, $f_2$, and $f_3$. After that, the features at different scales are preliminarily aligned by warping at the corresponding scales. The $f_1$ uses the optical flow at the original resolution, while $f_2$ and $f_3$ use the optical flow after 1x downsampling and 2x downsampling, respectively. After warping, the feature in the original resolution is further aligned by the align module according to the optical flow and the warping frame. Finally, the fusion module fuses three features in different scales to obtain the temporal contexts in three scales. 

For frames other than intra-coding frames, the context generation is divided into two branches: adjacent frame branch and key frame branch. \textbf{The adjacent frame branch} is almost identical to the intra-coding frame's context generation, with the only change being that the input changes from a reconstructed frame $\hat{x}_{t-1}$ to a decoder-side feature $F_{t-1}$. The architecture of the \textbf{key frame branch} is inherited from the adjacent frame branch, except that its input is the decoder-side feature of the key frame, and it will be warped $n$ times to be aligned from the key frame to the current frame ($n$ is the distance between the key frame and the current frame). The fusion module finally fuses the features obtained from two branches to obtain three temporal contexts in different scales.  

\vspace{-1.0em}
\subsection{Lookahead}
Since part of the quality structure of NVC consists of implicit quality losses in the lossy neural network parameters, to make the implicit quality structure more efficient, we try to introduce more information to the network that helps construct the quality structure. Therefore, we propose a lookahead strategy. It is designed to enable the encoder to learn a more forward-looking quality structure. Notably, to maintain the low delay setting, our approach incorporates only one future frame of lookahead information for the encoder. 

As illustrated in Fig.~\ref{fig-framework}, the lookahead module fuses the subsequent frame $x_{t+1}$ with the current temporal context through lookahead merging, producing the final encoder-side context $c_t^{enc}$. The decoder, however, utilizes the original context prior to lookahead fusion $c_t^{dec}$. 

\vspace{-0.8em}
\subsection{Layer-wise quantization scale with random quality training}
Due to insufficient training, a purely implicit quality structure is not efficient enough. Therefore, DCVC-DC~\cite{li2023neural} first introduces explicit hierarchical Lagrange multipliers in NVC to further optimize the quality structure. The Lagrangian weights are introduced to control the trade-off between bitrate and reconstruction quality in the training process. However, based on the experience in traditional video codecs, the design of the hierarchical quality structure includes not only the hierarchical Lagrangian multipliers controlling the RDO but also the hierarchical quantization parameters controlling the quantization. 

Therefore, we introduce a layer-wise quantization scale paired with hierarchical Lagrange multipliers to achieve a more stable and efficient quality structure jointly. In specific, we introduce different learnable layer-wise quantization scales in both encoder and decoder ($Q_{layer}^{enc}$ and $Q_{layer}^{dec}$ in Fig.~\ref{fig-framework}) for frames in different layers (frames using different hierarchical weights). 

Considering that the quality structure can not remain strictly stable during inference, we introduce a random quality training strategy to adapt the model to the fluctuations of the quality structure. In specific, the quantization scale of the first key frame (the frame with index 1) will be randomly scaled during training. That is, the quantization scales of the first key frame during training are calculated by:

\begin{equation}
\label{eq:randomQ}
\left\{
\begin{aligned}
Q_{enc} &= Q^{enc}_{layer} \times \omega \\
Q_{dec} &= Q^{dec}_{layer} \times \frac{1}{\omega}
\end{aligned}
\right.
\end{equation}

where the $Q_{enc}$ and the $Q_{dec}$ are the final quantization scale used in encoder and decoder, respectively. The $\omega$ is a random scale with values ranging from 0.8 to 1.2.

\vspace{-0.8em}
\section{Experiments}
\subsection{Experimental Settings}

\begin{table*}[t]
\caption{BD-Rate (\%) comparison for PSNR. The anchor is VTM-23.4 LDB. All codecs are under 96 frames with intra-period 32.\label{tab:performanceIP32}}
\centering
\begin{tabular}{cccccccc}
\toprule
& UVG & MCL-JCV & HEVC B & HEVC C & HEVC D & HEVC E & Average \\
\midrule
VTM-23.4    &0.00   &0.00   &0.00   &0.00   &0.00   &0.00   &0.00 \\
DCVC-TCM    &28.84  &41.39  &35.81  &71.65  &32.21  &91.01  &50.15  \\
DCVC-HEM    &-4.69  &5.94   &4.29   &28.13  &-3.30  &28.72  &9.85 \\
DCVC-DC     &-17.96 &-9.03  &-9.01  &-3.54  &\cellcolor{lightgray}\textbf{-24.99} &-10.29 &-12.47 \\
DCVC-FM     &-11.40 &-0.53  &-1.69  &4.63   &-19.40 &-8.58  &-6.16 \\
EHVC        &\cellcolor{lightgray}\textbf{-29.11} &\cellcolor{lightgray}\textbf{-15.12} &\cellcolor{lightgray}\textbf{-17.97} &\cellcolor{lightgray}\textbf{-5.96}  &-23.68 &\cellcolor{lightgray}\textbf{-10.98} &\cellcolor{lightgray}\textbf{-17.14} \\
\bottomrule
\end{tabular}
\vspace{-0.5em}
\end{table*}

\begin{figure*}[t]
  \centering
  \includegraphics[width=18cm]{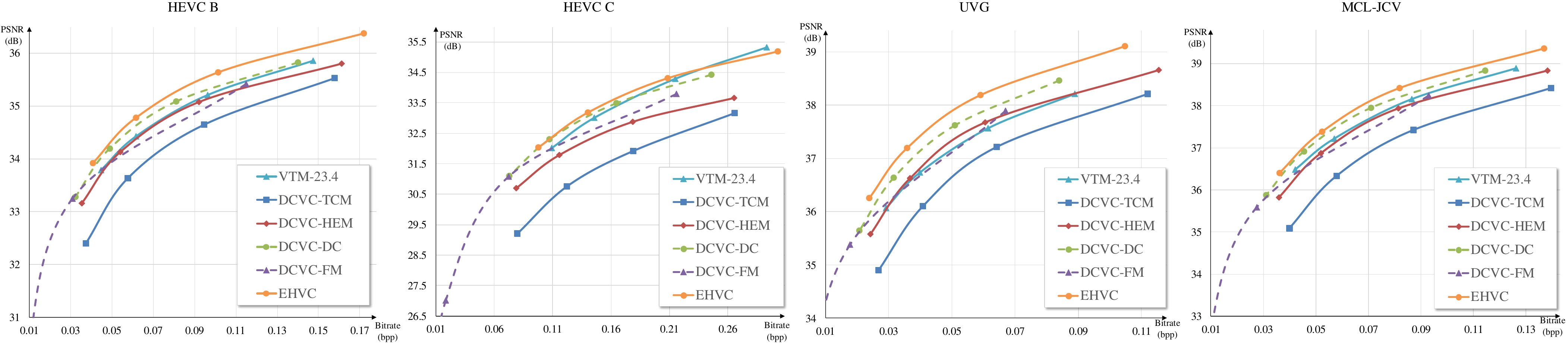}
  \caption{Rate-Distortion curves on HEVC B, HEVC C, UVG, and MCL-JCV dataset. The intra-period is 32 with 96 frames.}
  \label{fig-curve32}
  \vspace{-1.2em}
\end{figure*}

\subsubsection{Datasets}
To evaluate the effectiveness of our EHVC, we use six widely used and representative video compression datasets for testing: UVG~\cite{mercat2020uvg} and MCL-JCV~\cite{wang2016mcl}, HEVC~\cite{sullivan2012overview} Class B, C, D, and E. UVG and MCL-JCV are commonly used public video datasets for evaluating video compression algorithms, consisting of 7 and 30 high-quality videos with the resolution 1920$\times$1080, respectively. The HEVC dataset was selected by the video coding standardization organizations ISO/IEC and ITU-T to evaluate the performance of coding techniques. It contains video sequences in rich content with resolutions ranging from 416$\times$240 to 1920$\times$1080.

\vspace{-0.5em}
\subsubsection{Baseline Codecs}
To demonstrate the superiority of EHVC, we select the SOTA codecs in both traditional video codecs and NVCs as the baseline. For traditional codec, we utilize the VTM-23.4, a recent version of the reference software of the H.266/VVC standard~\cite{bross2021overview}. For NVCs, we utilize the DCVC family with SOTA performance, including DCVC-TCM~\cite{sheng2022temporal}, DCVC-HEM~\cite{li2022hybrid}, DCVC-DC~\cite{li2023neural}, and DCVC-FM~\cite{li2024neural}. In addition, for all the codecs, we test the performance under 96 frames with intra-period (IP) 32 and all frames with IP -1 (only the first frame is the intra-coding frame). Moreover, we measure the bitrate by bits-per-pixel (bpp) and assess the reconstruction quality by the peak signal-to-noise ratio (PSNR).

\vspace{-0.8em}
\subsection{Comparisons with Previous Methods}

\subsubsection{Rate-distortion Performance under Intra-period 32}
Table~\ref{tab:performanceIP32} shows the performance comparison under 96 frames with intra-period 32. Fig.~\ref{fig-curve32} shows the rate-distortion curves. As shown in the table, our EHVC achieves the best coding performance compared to the SOTA traditional video codec and the previous SOTA NVCs. Our EHVC achieves an average of \textbf{17.14\%} bitrate saving over VTM-23.4. In addition, our EHVC can save \textbf{10.98\%} more bitrate than previous SOTA NVC DCVC-FM over VTM-23.4.

\vspace{-0.5em}
\subsubsection{Rate-distortion Performance under Intra-period -1}

\begin{table*}[t]
\caption{BD-Rate (\%) comparison for PSNR. The anchor is VTM-23.4 LDB. All codecs are under all frames with intra-period -1.\label{tab:performanceIP-1}}
\centering
\begin{tabular}{cccccccc}
\toprule
& UVG & MCL-JCV & HEVC B & HEVC C & HEVC D & HEVC E & Average \\
\midrule
VTM-23.4    &0.00   &0.00   &0.00   &0.00   &0.00   &0.00   &0.00 \\
DCVC-TCM    &107.02 &77.26  &125.41 &143.62 &99.22  &544.82 &92.09 \\
DCVC-HEM    &46.42  &23.20  &42.46  &47.66  &19.55  &410.13 &98.24 \\
DCVC-DC     &7.62   &1.68   &10.21  &17.17  &-5.22  &119.52 &25.16 \\
DCVC-FM     &-8.30  &2.22   &1.00   &-2.14  &-17.03 &\cellcolor{lightgray}\textbf{-0.68}  &-4.16 \\
EHVC        &\cellcolor{lightgray}\textbf{-28.21} &\cellcolor{lightgray}\textbf{-15.50} &\cellcolor{lightgray}\textbf{-18.57} &\cellcolor{lightgray}\textbf{-16.85} &\cellcolor{lightgray}\textbf{-25.45} &2.32   &\cellcolor{lightgray}\textbf{-17.04} \\
\bottomrule
\end{tabular}
\end{table*}

\begin{figure*}[t]
  \centering
  \includegraphics[width=18cm]{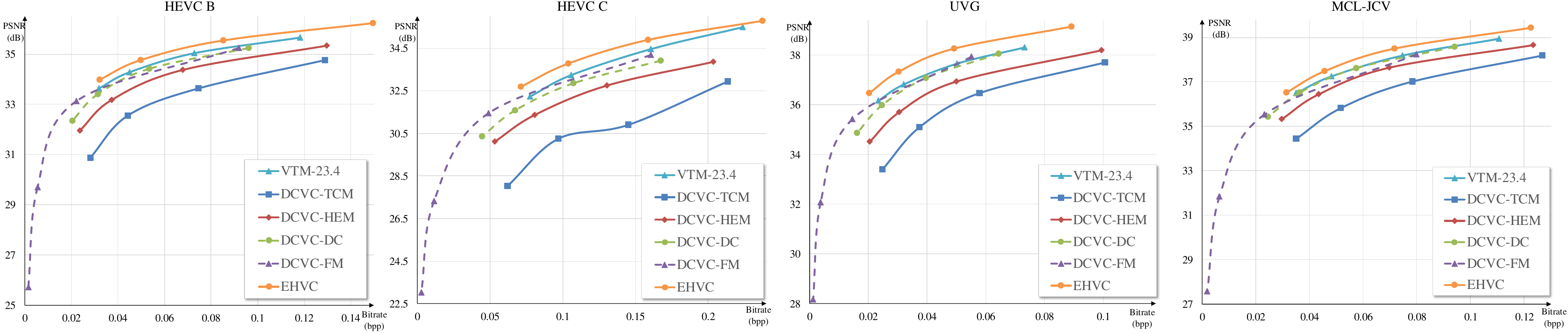}
  \caption{Rate-Distortion curves on HEVC B, HEVC C, UVG, and MCL-JCV dataset. The intra-period is -1 with all frames.}
  \label{fig-curve-1}
  \vspace{-1.2em}
\end{figure*}

As a more common and practical setting, intra-period -1 can better assess the stability of the reference and quality structures. The performance of codecs in all frames with intra-period -1 is tested and shown in table~\ref{tab:performanceIP-1}. Fig.~\ref{fig-curve-1} shows the rate-distortion curves. The table shows that our EHVC also achieves the best coding performance among SOTA video codecs. Our EHVC achieves an average of \textbf{17.04\%} bitrate saving over VTM-23.4. Moreover, it can save \textbf{12.88\%} more bitrate than previous SOTA NVC DCVC-FM over VTM-23.4.

\vspace{-0.5em}
\subsubsection{Performance of the quality structure}
As shown in Fig.~\ref{fig-framePSNR}, the quality structure of our EHVC is more stable than in the previous SOTA NVC DCVC-FM. Specifically, within each context generation refresh cycle (32 frames), the quality degradation of EHVC is about 1 dB, significantly smaller than the quality degradation of about 1.5 dB in DCVC-FM. 

Moreover, clear examples can be seen in the zoomed-in window of the figure. For example, for DCVC-FM, the quality of the 30th and 31st frames is almost the same, which does not conform to the designed hierarchical quality structure. In contrast, the quality of the 31st frame in EHVC is significantly higher than that of the 30th frame, which conforms to the design of the hierarchical quality structure. Another example is the 33rd frame and the 35th frame, where according to the designed hierarchical quality structure, frame 33 (using a weight of 1.2) should have higher quality than frame 35 (using a weight of 0.9). The performance of EHVC is consistent with the design, but frame 35 in DCVC-FM has a higher quality than frame 33 instead.

\vspace{-0.5em}
\subsubsection{Performance of the reference structure}
As shown in Fig.~\ref{fig-blackF}, the reference structure of our EHVC is also more stable than in the previous SOTA NVC DCVC-FM. Because of the extreme setting of the experiment and the fact that the model has not been exposed to such an extreme situation during the training stage, the reconstruction quality of the next frame of the no-information frame is still greatly affected in our EHVC. However, it is worth noting that the reconstruction quality of the subsequent frames quickly recovers and remains almost at the same level as the reconstruction quality before the no-information frame. This illustrates the stronger stability and robustness of the hierarchical reference structure matched to the hierarchical quality structure in EHVC.

\vspace{-0.8em}
\subsection{Complexity Analysis}
\begin{table}[t]
\caption{Complexity comparison among the proposed EHVC, DCVC-HEM, DCVC-DC, and DCVC-FM. The KMACs/pixel denotes the multiply-add operations per pixel. The $T_{enc}$ and $T_{dec}$ denote the average encoding and decoding time per frame, respectively.\label{tab:complexity}}
\centering
\begin{tabular}{cccc}
\toprule
NVC & kMACs/pixel & $T_{enc}$ (s) & $T_{dec}$ (s) \\
\midrule
DCVC-HEM~\cite{li2022hybrid}    & 1622  &0.82  &0.74  \\
DCVC-DC~\cite{li2023neural}     & 1344  &0.88  &0.79  \\
DCVC-FM~\cite{li2024neural}     & 1125  &0.91  &0.81  \\
EHVC                & 1311  &0.93  &0.84  \\
\bottomrule
\end{tabular}
\vspace{-1.8em}
\end{table}
We test the complexity of EHVC and compare it with the previous NVCs DCVC-HEM, DCVC-DC, and DCVC-FM. The results are shown in Table~\ref{tab:complexity}. From the results, the complexity of our EHVC is slightly higher than that of DCVC-FM. Considering that EHVC is built on top of DCVC-FM and that EHVC has a performance gain of over 10\% compared to DCVC-FM, the complexity introduced by EHVC's improvements is acceptable.

\vspace{-0.8em}
\subsection{Ablation Study}
\begin{table}[t]
\caption{Ablation study using BD-Rate (\%)\label{tab:ablation}}
\centering
\begin{tabular}{cccccc}
\toprule
& $M_a$ & $M_b$ & $M_c$ & $M_d$  & $M_e$\\
\midrule
\begin{tabular}[c]{@{}c@{}}Hierarchical \\ multi-reference\end{tabular} & 
 &\checkmark  &\checkmark &\checkmark &\checkmark  \\
 \midrule
Lookahead   &  &  &\checkmark &\checkmark &\checkmark   \\
\midrule
\begin{tabular}[c]{@{}c@{}}Layer-wise \\ quality scale\end{tabular}  &   &   &   &\checkmark &\checkmark   \\
\midrule
\begin{tabular}[c]{@{}c@{}}Random quality \\ training\end{tabular}   &   &   &   &    &\checkmark   \\
\midrule
BD-Rate(\%)  & 0.00&-10.47\% &-12.74\% &-13.09\% &-13.49\%  \\
\bottomrule
\end{tabular}
\vspace{-1.8em}
\end{table}

Table~\ref{tab:ablation} shows the ablation study on each improvement of EHVC. The BD-Rates in the table are the average BD-Rates on HEVC B, C, D, and E under all frames with intra-period -1. In this table, the baseline model $M_a$ is DCVC-FM. First, we test the performance of hierarchical multi-reference. This technique ($M_b$) achieves 10.47\% bitrate saving, demonstrating the benefits of aligning the reference structure with the quality structure. Then, integrating the lookahead strategy ($M_c$) raises the bitrate saving to 12.74\%. Moreover, the layer-wise quality scale ($M_d$) and the random quality training ($M_e$) further increase the bitrate saving to 13.09\% and 13.49\%, respectively.

\vspace{-0.8em}
\section{Conclusion}
In this paper, we analyze the problems in the hierarchical structure design of the previous SOTA NVCs: the unrefined design of the quality structure and the lack of a reference structure matching the quality structure. To address these challenges, we propose an efficient hierarchical neural video codec, EHVC. The hierarchical reference structure is first designed. To realize the reference structure, we design the multi-reference context generation module and finally eliminate the reference-quality mismatch. Then, we introduce the lookahead strategy to enhance the quality structure. In addition, we propose a layer-wise quality scale with random quality training to stabilize the quality structure. With these improvements, our EHVC achieves SOTA rate-distortion performance. 

However, despite the SOTA performance, there is still room for improvement in EHVC's hierarchical structure design inherited from DCVC-DC. How to design a more efficient and stable hierarchical structure is still the research direction of our future work.

\begin{acks}
This work was supported in part by the National Key Research and Development Plan under Grant 2024YFF0505702, and in part by the Natural Science Foundation of China under Grants 62171429 and 62021001.
\end{acks}

\
\bibliographystyle{ACM-Reference-Format}
\balance
\bibliography{sample-base}


\begin{thebibliography}{44}


\ifx \showCODEN    \undefined \def \showCODEN     #1{\unskip}     \fi
\ifx \showISBNx    \undefined \def \showISBNx     #1{\unskip}     \fi
\ifx \showISBNxiii \undefined \def \showISBNxiii  #1{\unskip}     \fi
\ifx \showISSN     \undefined \def \showISSN      #1{\unskip}     \fi
\ifx \showLCCN     \undefined \def \showLCCN      #1{\unskip}     \fi
\ifx \shownote     \undefined \def \shownote      #1{#1}          \fi
\ifx \showarticletitle \undefined \def \showarticletitle #1{#1}   \fi
\ifx \showURL      \undefined \def \showURL       {\relax}        \fi
\providecommand\bibfield[2]{#2}
\providecommand\bibinfo[2]{#2}
\providecommand\natexlab[1]{#1}
\providecommand\showeprint[2][]{arXiv:#2}

\bibitem[Agustsson et~al\mbox{.}(2020)]%
        {agustsson2020scale}
\bibfield{author}{\bibinfo{person}{Eirikur Agustsson}, \bibinfo{person}{David Minnen}, \bibinfo{person}{Nick Johnston}, \bibinfo{person}{Johannes Balle}, \bibinfo{person}{Sung~Jin Hwang}, {and} \bibinfo{person}{George Toderici}.} \bibinfo{year}{2020}\natexlab{}.
\newblock \showarticletitle{Scale-space flow for end-to-end optimized video compression}. In \bibinfo{booktitle}{\emph{Proceedings of the IEEE/CVF Conference on Computer Vision and Pattern Recognition}}. \bibinfo{pages}{8503--8512}.
\newblock


\bibitem[Alexandre et~al\mbox{.}(2023)]%
        {alexandre2023hierarchical}
\bibfield{author}{\bibinfo{person}{David Alexandre}, \bibinfo{person}{Hsueh-Ming Hang}, {and} \bibinfo{person}{Wen-Hsiao Peng}.} \bibinfo{year}{2023}\natexlab{}.
\newblock \showarticletitle{Hierarchical B-frame video coding using two-layer CANF without motion coding}. In \bibinfo{booktitle}{\emph{Proceedings of the IEEE/CVF Conference on Computer Vision and Pattern Recognition}}. \bibinfo{pages}{10249--10258}.
\newblock


\bibitem[Bian et~al\mbox{.}(2025)]%
        {bian2025augmented}
\bibfield{author}{\bibinfo{person}{Yifan Bian}, \bibinfo{person}{Chuanbo Tang}, \bibinfo{person}{Li Li}, {and} \bibinfo{person}{Dong Liu}.} \bibinfo{year}{2025}\natexlab{}.
\newblock \showarticletitle{Augmented Deep Contexts for Spatially Embedded Video Coding}. In \bibinfo{booktitle}{\emph{Proceedings of the Computer Vision and Pattern Recognition Conference}}. \bibinfo{pages}{2094--2104}.
\newblock


\bibitem[Bjontegaard(2001)]%
        {bjontegaard2001calculation}
\bibfield{author}{\bibinfo{person}{Gisle Bjontegaard}.} \bibinfo{year}{2001}\natexlab{}.
\newblock \showarticletitle{Calculation of average PSNR differences between RD-curves}.
\newblock \bibinfo{journal}{\emph{ITU SG16 Doc. VCEG-M33}} (\bibinfo{year}{2001}).
\newblock


\bibitem[Bossen et~al\mbox{.}(2021)]%
        {bossen2021vvc}
\bibfield{author}{\bibinfo{person}{Frank Bossen}, \bibinfo{person}{Karsten S{\"u}hring}, \bibinfo{person}{Adam Wieckowski}, {and} \bibinfo{person}{Shan Liu}.} \bibinfo{year}{2021}\natexlab{}.
\newblock \showarticletitle{VVC complexity and software implementation analysis}.
\newblock \bibinfo{journal}{\emph{IEEE Transactions on Circuits and Systems for Video Technology}} \bibinfo{volume}{31}, \bibinfo{number}{10} (\bibinfo{year}{2021}), \bibinfo{pages}{3765--3778}.
\newblock


\bibitem[Bross et~al\mbox{.}(2021a)]%
        {bross2021developments}
\bibfield{author}{\bibinfo{person}{Benjamin Bross}, \bibinfo{person}{Jianle Chen}, \bibinfo{person}{Jens-Rainer Ohm}, \bibinfo{person}{Gary~J Sullivan}, {and} \bibinfo{person}{Ye-Kui Wang}.} \bibinfo{year}{2021}\natexlab{a}.
\newblock \showarticletitle{Developments in international video coding standardization after AVC, with an overview of versatile video coding (VVC)}.
\newblock \bibinfo{journal}{\emph{Proc. IEEE}} \bibinfo{volume}{109}, \bibinfo{number}{9} (\bibinfo{year}{2021}), \bibinfo{pages}{1463--1493}.
\newblock


\bibitem[Bross et~al\mbox{.}(2021b)]%
        {bross2021overview}
\bibfield{author}{\bibinfo{person}{Benjamin Bross}, \bibinfo{person}{Ye-Kui Wang}, \bibinfo{person}{Yan Ye}, \bibinfo{person}{Shan Liu}, \bibinfo{person}{Jianle Chen}, \bibinfo{person}{Gary~J Sullivan}, {and} \bibinfo{person}{Jens-Rainer Ohm}.} \bibinfo{year}{2021}\natexlab{b}.
\newblock \showarticletitle{Overview of the versatile video coding (VVC) standard and its applications}.
\newblock \bibinfo{journal}{\emph{IEEE Transactions on Circuits and Systems for Video Technology}} \bibinfo{volume}{31}, \bibinfo{number}{10} (\bibinfo{year}{2021}), \bibinfo{pages}{3736--3764}.
\newblock


\bibitem[Chen et~al\mbox{.}(2023)]%
        {chen2023neural}
\bibfield{author}{\bibinfo{person}{Zhenghao Chen}, \bibinfo{person}{Lucas Relic}, \bibinfo{person}{Roberto Azevedo}, \bibinfo{person}{Yang Zhang}, \bibinfo{person}{Markus Gross}, \bibinfo{person}{Dong Xu}, \bibinfo{person}{Luping Zhou}, {and} \bibinfo{person}{Christopher Schroers}.} \bibinfo{year}{2023}\natexlab{}.
\newblock \showarticletitle{Neural video compression with spatio-temporal cross-covariance transformers}. In \bibinfo{booktitle}{\emph{Proceedings of the 31st ACM International Conference on Multimedia}}. \bibinfo{pages}{8543--8551}.
\newblock


\bibitem[Djelouah et~al\mbox{.}(2019)]%
        {djelouah2019neural}
\bibfield{author}{\bibinfo{person}{Abdelaziz Djelouah}, \bibinfo{person}{Joaquim Campos}, \bibinfo{person}{Simone Schaub-Meyer}, {and} \bibinfo{person}{Christopher Schroers}.} \bibinfo{year}{2019}\natexlab{}.
\newblock \showarticletitle{Neural inter-frame compression for video coding}. In \bibinfo{booktitle}{\emph{Proceedings of the IEEE/CVF international conference on computer vision}}. \bibinfo{pages}{6421--6429}.
\newblock


\bibitem[Girod et~al\mbox{.}(1995)]%
        {girod1995comparison}
\bibfield{author}{\bibinfo{person}{Bernd Girod}, \bibinfo{person}{Eckehard~G Steinbach}, {and} \bibinfo{person}{Niko Faerber}.} \bibinfo{year}{1995}\natexlab{}.
\newblock \showarticletitle{Comparison of the H. 263 and H. 261 video compression standards}. In \bibinfo{booktitle}{\emph{Standards and Common Interfaces for Video Information Systems: A Critical Review}}, Vol.~\bibinfo{volume}{10282}. SPIE, \bibinfo{pages}{230--248}.
\newblock


\bibitem[Guo et~al\mbox{.}(2023)]%
        {guo2023enhanced}
\bibfield{author}{\bibinfo{person}{Haifeng Guo}, \bibinfo{person}{Sam Kwong}, \bibinfo{person}{Dongjie Ye}, {and} \bibinfo{person}{Shiqi Wang}.} \bibinfo{year}{2023}\natexlab{}.
\newblock \showarticletitle{Enhanced context mining and filtering for learned video compression}.
\newblock \bibinfo{journal}{\emph{IEEE Transactions on Multimedia}}  \bibinfo{volume}{26} (\bibinfo{year}{2023}), \bibinfo{pages}{3814--3826}.
\newblock


\bibitem[Habibian et~al\mbox{.}(2019)]%
        {habibian2019video}
\bibfield{author}{\bibinfo{person}{Amirhossein Habibian}, \bibinfo{person}{Ties~van Rozendaal}, \bibinfo{person}{Jakub~M Tomczak}, {and} \bibinfo{person}{Taco~S Cohen}.} \bibinfo{year}{2019}\natexlab{}.
\newblock \showarticletitle{Video compression with rate-distortion autoencoders}. In \bibinfo{booktitle}{\emph{Proceedings of the IEEE/CVF international conference on computer vision}}. \bibinfo{pages}{7033--7042}.
\newblock


\bibitem[Ho et~al\mbox{.}(2022)]%
        {ho2022canf}
\bibfield{author}{\bibinfo{person}{Yung-Han Ho}, \bibinfo{person}{Chih-Peng Chang}, \bibinfo{person}{Peng-Yu Chen}, \bibinfo{person}{Alessandro Gnutti}, {and} \bibinfo{person}{Wen-Hsiao Peng}.} \bibinfo{year}{2022}\natexlab{}.
\newblock \showarticletitle{Canf-vc: Conditional augmented normalizing flows for video compression}. In \bibinfo{booktitle}{\emph{European Conference on Computer Vision}}. Springer, \bibinfo{pages}{207--223}.
\newblock


\bibitem[Hu et~al\mbox{.}(2020)]%
        {hu2020improving}
\bibfield{author}{\bibinfo{person}{Zhihao Hu}, \bibinfo{person}{Zhenghao Chen}, \bibinfo{person}{Dong Xu}, \bibinfo{person}{Guo Lu}, \bibinfo{person}{Wanli Ouyang}, {and} \bibinfo{person}{Shuhang Gu}.} \bibinfo{year}{2020}\natexlab{}.
\newblock \showarticletitle{Improving deep video compression by resolution-adaptive flow coding}. In \bibinfo{booktitle}{\emph{Computer Vision--ECCV 2020: 16th European Conference, Glasgow, UK, August 23--28, 2020, Proceedings, Part II 16}}. Springer, \bibinfo{pages}{193--209}.
\newblock


\bibitem[Hu et~al\mbox{.}(2022)]%
        {hu2022coarse}
\bibfield{author}{\bibinfo{person}{Zhihao Hu}, \bibinfo{person}{Guo Lu}, \bibinfo{person}{Jinyang Guo}, \bibinfo{person}{Shan Liu}, \bibinfo{person}{Wei Jiang}, {and} \bibinfo{person}{Dong Xu}.} \bibinfo{year}{2022}\natexlab{}.
\newblock \showarticletitle{Coarse-to-fine deep video coding with hyperprior-guided mode prediction}. In \bibinfo{booktitle}{\emph{Proceedings of the IEEE/CVF Conference on Computer Vision and Pattern Recognition}}. \bibinfo{pages}{5921--5930}.
\newblock


\bibitem[Ladune et~al\mbox{.}(2021)]%
        {ladune2021conditional}
\bibfield{author}{\bibinfo{person}{Th{\'e}o Ladune}, \bibinfo{person}{Pierrick Philippe}, \bibinfo{person}{Wassim Hamidouche}, \bibinfo{person}{Lu Zhang}, {and} \bibinfo{person}{Olivier D{\'e}forges}.} \bibinfo{year}{2021}\natexlab{}.
\newblock \showarticletitle{Conditional coding for flexible learned video compression}.
\newblock \bibinfo{journal}{\emph{arXiv preprint arXiv:2104.07930}} (\bibinfo{year}{2021}).
\newblock


\bibitem[Li et~al\mbox{.}(2021)]%
        {li2021deep}
\bibfield{author}{\bibinfo{person}{Jiahao Li}, \bibinfo{person}{Bin Li}, {and} \bibinfo{person}{Yan Lu}.} \bibinfo{year}{2021}\natexlab{}.
\newblock \showarticletitle{Deep contextual video compression}.
\newblock \bibinfo{journal}{\emph{Advances in Neural Information Processing Systems}}  \bibinfo{volume}{34} (\bibinfo{year}{2021}), \bibinfo{pages}{18114--18125}.
\newblock


\bibitem[Li et~al\mbox{.}(2022)]%
        {li2022hybrid}
\bibfield{author}{\bibinfo{person}{Jiahao Li}, \bibinfo{person}{Bin Li}, {and} \bibinfo{person}{Yan Lu}.} \bibinfo{year}{2022}\natexlab{}.
\newblock \showarticletitle{Hybrid spatial-temporal entropy modelling for neural video compression}. In \bibinfo{booktitle}{\emph{Proceedings of the 30th ACM International Conference on Multimedia}}. \bibinfo{pages}{1503--1511}.
\newblock


\bibitem[Li et~al\mbox{.}(2023)]%
        {li2023neural}
\bibfield{author}{\bibinfo{person}{Jiahao Li}, \bibinfo{person}{Bin Li}, {and} \bibinfo{person}{Yan Lu}.} \bibinfo{year}{2023}\natexlab{}.
\newblock \showarticletitle{Neural video compression with diverse contexts}. In \bibinfo{booktitle}{\emph{Proceedings of the IEEE/CVF conference on computer vision and pattern recognition}}. \bibinfo{pages}{22616--22626}.
\newblock


\bibitem[Li et~al\mbox{.}(2024)]%
        {li2024neural}
\bibfield{author}{\bibinfo{person}{Jiahao Li}, \bibinfo{person}{Bin Li}, {and} \bibinfo{person}{Yan Lu}.} \bibinfo{year}{2024}\natexlab{}.
\newblock \showarticletitle{Neural video compression with feature modulation}. In \bibinfo{booktitle}{\emph{Proceedings of the IEEE/CVF Conference on Computer Vision and Pattern Recognition}}. \bibinfo{pages}{26099--26108}.
\newblock


\bibitem[Liao et~al\mbox{.}(2025)]%
        {liao2025efvc}
\bibfield{author}{\bibinfo{person}{Junqi Liao}, \bibinfo{person}{Li Li}, \bibinfo{person}{Dong Liu}, {and} \bibinfo{person}{Houqiang Li}.} \bibinfo{year}{2025}\natexlab{}.
\newblock \showarticletitle{EFVC: Error-Propagation-Free Neural Video Coding with Reversible Transform}. In \bibinfo{booktitle}{\emph{2025 IEEE International Symposium on Circuits and Systems (ISCAS)}}. IEEE, \bibinfo{pages}{1--5}.
\newblock


\bibitem[Lin et~al\mbox{.}(2020)]%
        {lin2020m}
\bibfield{author}{\bibinfo{person}{Jianping Lin}, \bibinfo{person}{Dong Liu}, \bibinfo{person}{Houqiang Li}, {and} \bibinfo{person}{Feng Wu}.} \bibinfo{year}{2020}\natexlab{}.
\newblock \showarticletitle{M-LVC: Multiple frames prediction for learned video compression}. In \bibinfo{booktitle}{\emph{Proceedings of the IEEE/CVF conference on computer vision and pattern recognition}}. \bibinfo{pages}{3546--3554}.
\newblock


\bibitem[Liu et~al\mbox{.}(2020a)]%
        {liu2020neural}
\bibfield{author}{\bibinfo{person}{Haojie Liu}, \bibinfo{person}{Ming Lu}, \bibinfo{person}{Zhan Ma}, \bibinfo{person}{Fan Wang}, \bibinfo{person}{Zhihuang Xie}, \bibinfo{person}{Xun Cao}, {and} \bibinfo{person}{Yao Wang}.} \bibinfo{year}{2020}\natexlab{a}.
\newblock \showarticletitle{Neural video coding using multiscale motion compensation and spatiotemporal context model}.
\newblock \bibinfo{journal}{\emph{IEEE Transactions on Circuits and Systems for Video Technology}} \bibinfo{volume}{31}, \bibinfo{number}{8} (\bibinfo{year}{2020}), \bibinfo{pages}{3182--3196}.
\newblock


\bibitem[Liu et~al\mbox{.}(2020b)]%
        {liu2020conditional}
\bibfield{author}{\bibinfo{person}{Jerry Liu}, \bibinfo{person}{Shenlong Wang}, \bibinfo{person}{Wei-Chiu Ma}, \bibinfo{person}{Meet Shah}, \bibinfo{person}{Rui Hu}, \bibinfo{person}{Pranaab Dhawan}, {and} \bibinfo{person}{Raquel Urtasun}.} \bibinfo{year}{2020}\natexlab{b}.
\newblock \showarticletitle{Conditional entropy coding for efficient video compression}. In \bibinfo{booktitle}{\emph{European Conference on Computer Vision}}. Springer, \bibinfo{pages}{453--468}.
\newblock


\bibitem[Lombardo et~al\mbox{.}(2019)]%
        {lombardo2019deep}
\bibfield{author}{\bibinfo{person}{Salvator Lombardo}, \bibinfo{person}{Jun Han}, \bibinfo{person}{Christopher Schroers}, {and} \bibinfo{person}{Stephan Mandt}.} \bibinfo{year}{2019}\natexlab{}.
\newblock \showarticletitle{Deep generative video compression}.
\newblock \bibinfo{journal}{\emph{Advances in Neural Information Processing Systems}}  \bibinfo{volume}{32} (\bibinfo{year}{2019}).
\newblock


\bibitem[Lu et~al\mbox{.}(2019)]%
        {lu2019dvc}
\bibfield{author}{\bibinfo{person}{Guo Lu}, \bibinfo{person}{Wanli Ouyang}, \bibinfo{person}{Dong Xu}, \bibinfo{person}{Xiaoyun Zhang}, \bibinfo{person}{Chunlei Cai}, {and} \bibinfo{person}{Zhiyong Gao}.} \bibinfo{year}{2019}\natexlab{}.
\newblock \showarticletitle{Dvc: An end-to-end deep video compression framework}. In \bibinfo{booktitle}{\emph{Proceedings of the IEEE/CVF conference on computer vision and pattern recognition}}. \bibinfo{pages}{11006--11015}.
\newblock


\bibitem[Lu et~al\mbox{.}(2020)]%
        {lu2020end}
\bibfield{author}{\bibinfo{person}{Guo Lu}, \bibinfo{person}{Xiaoyun Zhang}, \bibinfo{person}{Wanli Ouyang}, \bibinfo{person}{Li Chen}, \bibinfo{person}{Zhiyong Gao}, {and} \bibinfo{person}{Dong Xu}.} \bibinfo{year}{2020}\natexlab{}.
\newblock \showarticletitle{An end-to-end learning framework for video compression}.
\newblock \bibinfo{journal}{\emph{IEEE transactions on pattern analysis and machine intelligence}} \bibinfo{volume}{43}, \bibinfo{number}{10} (\bibinfo{year}{2020}), \bibinfo{pages}{3292--3308}.
\newblock


\bibitem[Mentzer et~al\mbox{.}(2022)]%
        {mentzer2022vct}
\bibfield{author}{\bibinfo{person}{Fabian Mentzer}, \bibinfo{person}{George Toderici}, \bibinfo{person}{David Minnen}, \bibinfo{person}{Sung-Jin Hwang}, \bibinfo{person}{Sergi Caelles}, \bibinfo{person}{Mario Lucic}, {and} \bibinfo{person}{Eirikur Agustsson}.} \bibinfo{year}{2022}\natexlab{}.
\newblock \showarticletitle{VCT: A video compression transformer}.
\newblock \bibinfo{journal}{\emph{arXiv preprint arXiv:2206.07307}} (\bibinfo{year}{2022}).
\newblock


\bibitem[Mercat et~al\mbox{.}(2021)]%
        {mercat2021comparative}
\bibfield{author}{\bibinfo{person}{Alexandre Mercat}, \bibinfo{person}{Arttu M{\"a}kinen}, \bibinfo{person}{Joose Sainio}, \bibinfo{person}{Ari Lemmetti}, \bibinfo{person}{Marko Viitanen}, {and} \bibinfo{person}{Jarno Vanne}.} \bibinfo{year}{2021}\natexlab{}.
\newblock \showarticletitle{Comparative rate-distortion-complexity analysis of VVC and HEVC video codecs}.
\newblock \bibinfo{journal}{\emph{IEEE Access}}  \bibinfo{volume}{9} (\bibinfo{year}{2021}), \bibinfo{pages}{67813--67828}.
\newblock


\bibitem[Mercat et~al\mbox{.}(2020)]%
        {mercat2020uvg}
\bibfield{author}{\bibinfo{person}{Alexandre Mercat}, \bibinfo{person}{Marko Viitanen}, {and} \bibinfo{person}{Jarno Vanne}.} \bibinfo{year}{2020}\natexlab{}.
\newblock \showarticletitle{UVG dataset: 50/120fps 4K sequences for video codec analysis and development}. In \bibinfo{booktitle}{\emph{Proceedings of the 11th ACM multimedia systems conference}}. \bibinfo{pages}{297--302}.
\newblock


\bibitem[Park and Kim(2019)]%
        {park2019deep}
\bibfield{author}{\bibinfo{person}{Woonsung Park} {and} \bibinfo{person}{Munchurl Kim}.} \bibinfo{year}{2019}\natexlab{}.
\newblock \showarticletitle{Deep predictive video compression with bi-directional prediction}.
\newblock \bibinfo{journal}{\emph{arXiv preprint arXiv:1904.02909}} (\bibinfo{year}{2019}).
\newblock


\bibitem[Park and Kim(2020)]%
        {park2020deep}
\bibfield{author}{\bibinfo{person}{Woonsung Park} {and} \bibinfo{person}{Munchurl Kim}.} \bibinfo{year}{2020}\natexlab{}.
\newblock \showarticletitle{Deep predictive video compression using mode-selective uni-and bi-directional predictions based on multi-frame hypothesis}.
\newblock \bibinfo{journal}{\emph{IEEE Access}}  \bibinfo{volume}{9} (\bibinfo{year}{2020}), \bibinfo{pages}{72--85}.
\newblock


\bibitem[Qi et~al\mbox{.}(2023)]%
        {qi2023motion}
\bibfield{author}{\bibinfo{person}{Linfeng Qi}, \bibinfo{person}{Jiahao Li}, \bibinfo{person}{Bin Li}, \bibinfo{person}{Houqiang Li}, {and} \bibinfo{person}{Yan Lu}.} \bibinfo{year}{2023}\natexlab{}.
\newblock \showarticletitle{Motion information propagation for neural video compression}. In \bibinfo{booktitle}{\emph{Proceedings of the IEEE/CVF Conference on Computer Vision and Pattern Recognition}}. \bibinfo{pages}{6111--6120}.
\newblock


\bibitem[Rippel et~al\mbox{.}(2021)]%
        {rippel2021elf}
\bibfield{author}{\bibinfo{person}{Oren Rippel}, \bibinfo{person}{Alexander~G Anderson}, \bibinfo{person}{Kedar Tatwawadi}, \bibinfo{person}{Sanjay Nair}, \bibinfo{person}{Craig Lytle}, {and} \bibinfo{person}{Lubomir Bourdev}.} \bibinfo{year}{2021}\natexlab{}.
\newblock \showarticletitle{Elf-vc: Efficient learned flexible-rate video coding}. In \bibinfo{booktitle}{\emph{Proceedings of the IEEE/CVF International Conference on Computer Vision}}. \bibinfo{pages}{14479--14488}.
\newblock


\bibitem[Sheng et~al\mbox{.}(2022)]%
        {sheng2022temporal}
\bibfield{author}{\bibinfo{person}{Xihua Sheng}, \bibinfo{person}{Jiahao Li}, \bibinfo{person}{Bin Li}, \bibinfo{person}{Li Li}, \bibinfo{person}{Dong Liu}, {and} \bibinfo{person}{Yan Lu}.} \bibinfo{year}{2022}\natexlab{}.
\newblock \showarticletitle{Temporal context mining for learned video compression}.
\newblock \bibinfo{journal}{\emph{IEEE Transactions on Multimedia}}  \bibinfo{volume}{25} (\bibinfo{year}{2022}), \bibinfo{pages}{7311--7322}.
\newblock


\bibitem[Sheng et~al\mbox{.}(2025)]%
        {sheng2025bi}
\bibfield{author}{\bibinfo{person}{Xihua Sheng}, \bibinfo{person}{Li Li}, \bibinfo{person}{Dong Liu}, {and} \bibinfo{person}{Shiqi Wang}.} \bibinfo{year}{2025}\natexlab{}.
\newblock \showarticletitle{Bi-Directional Deep Contextual Video Compression}.
\newblock \bibinfo{journal}{\emph{IEEE Transactions on Multimedia}} (\bibinfo{year}{2025}).
\newblock


\bibitem[Sullivan et~al\mbox{.}(2012)]%
        {sullivan2012overview}
\bibfield{author}{\bibinfo{person}{Gary~J Sullivan}, \bibinfo{person}{Jens-Rainer Ohm}, \bibinfo{person}{Woo-Jin Han}, {and} \bibinfo{person}{Thomas Wiegand}.} \bibinfo{year}{2012}\natexlab{}.
\newblock \showarticletitle{Overview of the high efficiency video coding (HEVC) standard}.
\newblock \bibinfo{journal}{\emph{IEEE Transactions on circuits and systems for video technology}} \bibinfo{volume}{22}, \bibinfo{number}{12} (\bibinfo{year}{2012}), \bibinfo{pages}{1649--1668}.
\newblock


\bibitem[Tang et~al\mbox{.}(2025)]%
        {tang2025neural}
\bibfield{author}{\bibinfo{person}{Chuanbo Tang}, \bibinfo{person}{Zhuoyuan Li}, \bibinfo{person}{Yifan Bian}, \bibinfo{person}{Li Li}, {and} \bibinfo{person}{Dong Liu}.} \bibinfo{year}{2025}\natexlab{}.
\newblock \showarticletitle{Neural Video Compression with Context Modulation}. In \bibinfo{booktitle}{\emph{Proceedings of the Computer Vision and Pattern Recognition Conference}}. \bibinfo{pages}{12553--12563}.
\newblock


\bibitem[Wang et~al\mbox{.}(2016)]%
        {wang2016mcl}
\bibfield{author}{\bibinfo{person}{Haiqiang Wang}, \bibinfo{person}{Weihao Gan}, \bibinfo{person}{Sudeng Hu}, \bibinfo{person}{Joe~Yuchieh Lin}, \bibinfo{person}{Lina Jin}, \bibinfo{person}{Longguang Song}, \bibinfo{person}{Ping Wang}, \bibinfo{person}{Ioannis Katsavounidis}, \bibinfo{person}{Anne Aaron}, {and} \bibinfo{person}{C-C~Jay Kuo}.} \bibinfo{year}{2016}\natexlab{}.
\newblock \showarticletitle{MCL-JCV: a JND-based H. 264/AVC video quality assessment dataset}. In \bibinfo{booktitle}{\emph{2016 IEEE international conference on image processing (ICIP)}}. IEEE, \bibinfo{pages}{1509--1513}.
\newblock


\bibitem[Wang et~al\mbox{.}(2021)]%
        {wang2021high}
\bibfield{author}{\bibinfo{person}{Ye-Kui Wang}, \bibinfo{person}{Robert Skupin}, \bibinfo{person}{Miska~M Hannuksela}, \bibinfo{person}{Sachin Deshpande}, \bibinfo{person}{Virginie Drugeon}, \bibinfo{person}{Rickard Sj{\"o}berg}, \bibinfo{person}{Byeongdoo Choi}, \bibinfo{person}{Vadim Seregin}, \bibinfo{person}{Yago Sanchez}, \bibinfo{person}{Jill~M Boyce}, {et~al\mbox{.}}} \bibinfo{year}{2021}\natexlab{}.
\newblock \showarticletitle{The high-level syntax of the versatile video coding (VVC) standard}.
\newblock \bibinfo{journal}{\emph{IEEE Transactions on Circuits and Systems for Video Technology}} \bibinfo{volume}{31}, \bibinfo{number}{10} (\bibinfo{year}{2021}), \bibinfo{pages}{3779--3800}.
\newblock


\bibitem[Xiang et~al\mbox{.}(2022)]%
        {xiang2022mimt}
\bibfield{author}{\bibinfo{person}{Jinxi Xiang}, \bibinfo{person}{Kuan Tian}, {and} \bibinfo{person}{Jun Zhang}.} \bibinfo{year}{2022}\natexlab{}.
\newblock \showarticletitle{Mimt: Masked image modeling transformer for video compression}. In \bibinfo{booktitle}{\emph{The Eleventh International Conference on Learning Representations}}.
\newblock


\bibitem[Yang et~al\mbox{.}(2020)]%
        {yang2020learning}
\bibfield{author}{\bibinfo{person}{Ren Yang}, \bibinfo{person}{Fabian Mentzer}, \bibinfo{person}{Luc~Van Gool}, {and} \bibinfo{person}{Radu Timofte}.} \bibinfo{year}{2020}\natexlab{}.
\newblock \showarticletitle{Learning for video compression with hierarchical quality and recurrent enhancement}. In \bibinfo{booktitle}{\emph{Proceedings of the IEEE/CVF Conference on Computer Vision and Pattern Recognition}}. \bibinfo{pages}{6628--6637}.
\newblock


\bibitem[Yilmaz and Tekalp(2020)]%
        {yilmaz2020end}
\bibfield{author}{\bibinfo{person}{M~Akin Yilmaz} {and} \bibinfo{person}{A~Murat Tekalp}.} \bibinfo{year}{2020}\natexlab{}.
\newblock \showarticletitle{End-to-end rate-distortion optimization for bi-directional learned video compression}. In \bibinfo{booktitle}{\emph{2020 IEEE International Conference on Image Processing (ICIP)}}. IEEE, \bibinfo{pages}{1311--1315}.
\newblock


\bibitem[Zhang et~al\mbox{.}(2023)]%
        {zhang2023neural}
\bibfield{author}{\bibinfo{person}{Yiwei Zhang}, \bibinfo{person}{Guo Lu}, \bibinfo{person}{Yunuo Chen}, \bibinfo{person}{Shen Wang}, \bibinfo{person}{Yibo Shi}, \bibinfo{person}{Jing Wang}, {and} \bibinfo{person}{Li Song}.} \bibinfo{year}{2023}\natexlab{}.
\newblock \showarticletitle{Neural rate control for learned video compression}. In \bibinfo{booktitle}{\emph{The Twelfth International Conference on Learning Representations}}.
\newblock


\end{thebibliography}

\end{document}